**The effects of ionic liquids on the thermodynamics of $H_2$ activation by frustrated Lewis pairs: a density functional theory study**

Xiaoqing Liu,[a†*] Xue Li,[†a] Xiaoqian Yao, Weizhen Zhao,[b] Lei Liu[b*]

[a] School of Chemical and Environmental Engineering, North University of China, Taiyuan, 030051, China

[b] Beijing Key Laboratory of Ionic Liquids Clean Process, Key Laboratory of Green Process and Engineering, Institute of Process Engineering, Chinese Academy of Sciences, Beijing 100049, China

Corresponding authors:
Xiaoqing Liu, xiaoqing@nuc.edu.cn;
Lei Liu, liulei@ipe.ac.cn; liulei3039@gmail.com;

† These authors contributed equally to the work

**Abstract:** Nowadays, hydrogen activation by frustrated Lewis pairs (FLPs) and their applications have been demonstrated to be one of emerge research topics in the field of catalysis. Previous studies have shown that the thermodynamics of these reaction is determined by electronic structures of FLPs and solvents. Herein, we investigated the systems consisting of typical FLPs and ionic liquids (ILs), which are well known by their large number of types and excellent solvent effects. The density functional theory (DFT) calculations were performed to study the thermodynamics for $H_2$ activation by both inter- and intra-molecular FLPs, as well as the individual components. The results show that the computed overall Gibbs free energies in ILs are more negative than that computed in toluene. Through the thermodynamics partitioning, we find that ILs favor the H-H cleavage elemental step, while disfavored the elemental steps of proton attachment, hydride attachment and zwitterionic stabilization. Moreover, the results show that these effects are strongly dependent on the type of FLPs, where intra-molecular FLPs are more effected compared to the inter-molecular FLPs.

**Introduction**

Almost two decades ago, Stephan and his group reported that a covalently linked phosphino–borane molecule, $Mes_2P(C_6F_4)B(C_6F_5)_2$, of which no classic Lewis-base adducts (i.e. dimers) have been found in the toluene solution. Instead, the nuclear magnetic resonance (NMR) spectroscopy confirmed that such molecule existed as a monomer. Interestingly, the authors found that the system was able to reversibly split hydrogen molecules ($H_2$) at temperatures ranging from 25 °C to 80 °C and under the pressure being 1 atm. They assumed that the $H_2$ activation is because of the bulky hindrance of MesP- and $C_6F_5$- groups, and named such systems as frustrated Lewis pairs (FLPs)[1]. After that, the concept of FLPs have been largely extended[2-3], mainly by two groups, i.e. Stephan's group focuses on the inter-molecular FLPs[4] and Erker's group focuses on the intra-molecular FLPs [5]. Nowadays the FLPs chemistry has become one of the important research topic in the field of catalysis and has been shown many important applications[6-8], such as capture and activation a series of small molecules, i.e. $H_2$, $CO_2$, $SO_2$[9-11] and hydrogenation reactions, including reduction of $CO_2$ to chemical products[12], and saturation of ethylene[13], acrylic[14],6,6-dimethylpentafulvene[15] and complexation of nitrous oxide[16].

Theoretical studies revealed that the reactivity of FLPs is strongly determined by the geometric parameters of Lewis acid (LA) and Lewis base (LB) complexes[17-18]. Taken the $H_2$ activation as an example, previous density functional theory (DFT) calculations showed that the suitable distances between the acid and base centers are from 3 Å to 5 Å[19]. Further computational analysis found that the cooperation strength of the LA and LB is the second main factor to control the reactivity between FLPs and $H_2$. The modifications of substitute groups have been shown to turn the thermodynamic behaviors of $H_2$ activation by FLPs[19], as well as their kinetics[20]. On the other hand, it was commonly shown that one of the important contributions to the thermodynamics is the solvent effect (i.e. solvation Gibbs free energies). Based on the COSMO-RS (conductor-like screening model for realistic solvents) calculations, Grimme et al. showed that solvent contributions are around -10 kcal mol$^{-1}$ [21]. The straightforward method is to change the type of solvents, where ionic liquids (ILs), excellent green solvents, had been proposed and examined as the reaction media for the $H_2$ activation by FLPs[22].

Recently, Brown et al reported a spectroscopic study on the structures of the typical FLP, $t$Bu$_3$P/B(C$_6$F$_5$)$_3$, in a IL of C$_{10min}$NTf$_2$[23]. The author found that the amount of effective FLPs complexes had been increased to be ca. 20% in ILs, compared that in toluene (which is only ca. 2 %[24]). The results indicates that ILs could largely improve the catalytic performance of FLPs, i.e. $H_2$ activation, which was also examined in their study. Inspired by that work, we selected eighteen inter- and intra-molecular FLPs, and investigated their thermodynamic properties of $H_2$ activation in toluene and ILs by DFT calculations. We found that the $H_2$ activation reactions are generally more favored in ILs compared to the traditional organic solvents (i.e. toluene). The computed overall Gibbs free energies (ΔG) in ILs are more negative than that in toluene by ca. -10 kcal mol$^{-1}$, which are also determined by the types of FLPs. Hence, we believe that that ILs could improve the catalytic performance of FLPs for $H_2$ activation, and most likely for other important applications of FLPs.

**Computational details**

All DFT calculations were performed by employing the G09 suit of program[25]. The geometries of all species discussed in this work have been optimized at the B3LYP/6-31G* level of theory[26-27]. The B3LYP hybrid functional have been shown to give accurate geometries and energies for FLP chemistry by previous computational studies[28-30], and previous benchmark calculations[19, 31] revealed that the computed ΔG for FLP reactions are somehow independent on the selected functional (i.e. B97D, M062X) and basis set (i.e. 6-31G* and 6-31+G**). Moreover, the adopted computational approach is not to produce accurate Gibbs energies for direct comparison with the experimentally measured quantities, but to provide qualitative trends for predicting FLP's thermochemistry. For each located stationary point, we performed a vibration analysis on the optimized structure at the same level of theory to confirm as a true minimum[29]. In terms of solvent effects, the SMD method[32] in the SCRF framework was used in the DFT calculations. For toluene, a dielectric constant, $\varepsilon$, of 2.37 was used, while the value of $\varepsilon$ was set to be 24.85, according to previous studies[22, 33]. Lastly, the natural bond orbital (NBO)[34] analysis were performed at the B3LYP/6-311++G** level of theory through single point calculations to obtain the natural population analysis (NPA) charges and dipole moments for the zwitterionic products. The selected experimental repoted FLPs are sumarized in **Scheme 1**[1-2, 14, 35-43], containging: 1) both inter- and intra-molecular FLPs; 2) typical Lewis acids (Boron, B, contained molecules), and Lewis bases

(Phousporus/Nitrigen, P/N, contained molecules); and 3) H$_2$ activations, and no reactions.

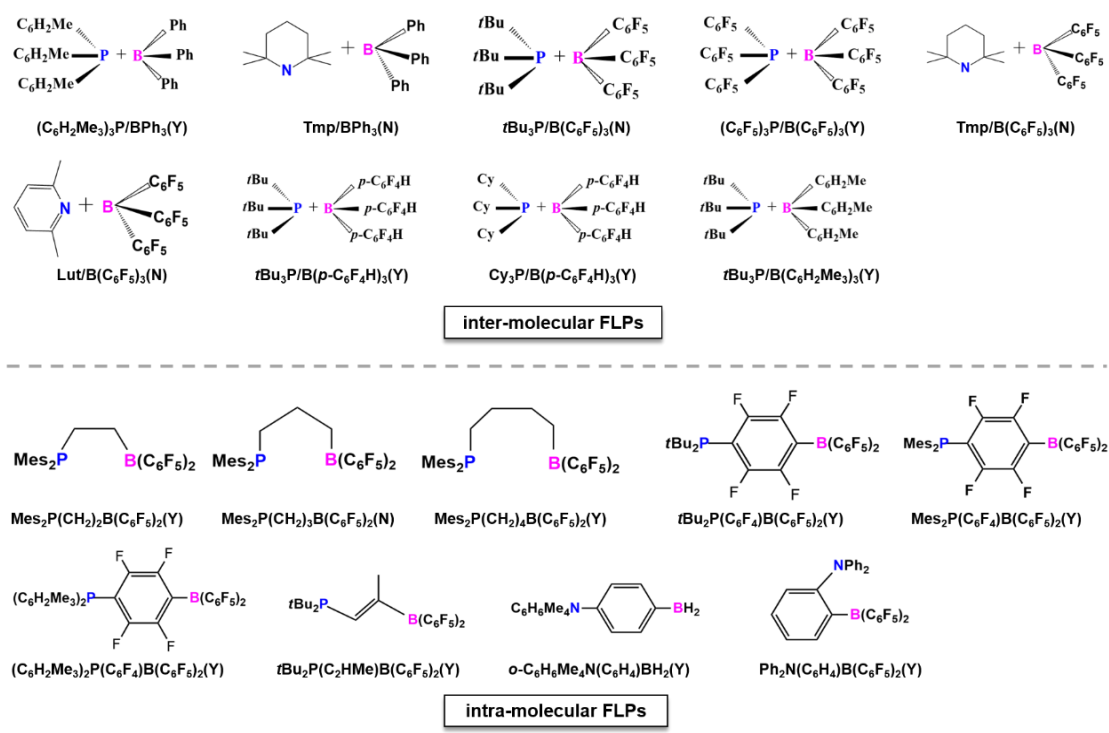

**Scheme 1**. Investigated FLPs in this work. Notation Y and N stand for experimentally reported H$_2$ activation and no reaction, respectively.

**Results and Discussion**

The computed ΔG of H$_2$ activation by investigated FLPs in gas phase, toluene and ILs are provided in **Figure 1** (for the reaction of LA + LB + H$_2$ → [LAH$^-$][LBH$^+$]). At first, we compare the computed ΔG in toluene (blue bars) with the experimental observations on the reactivity of FLPs with H$_2$ in toluene solutions. Generally, the systems with computed large positive ΔG values (i.e. > + 10 kcal mol$^{-1}$) are inactive with H$_2$, while the ones with slightly positive (i.e. < + 5 kcal mol$^{-1}$) or negative ΔG values are able to active H$_2$ and form zwitterionic products. Overall, the computed ΔG are in good agreement with the experimental observations, and previous theoretical calculations[29]. For instance, large positive values have been found for the unreactive FLPs, such as (C$_6$H$_2$Me$_3$)$_3$P/BPh$_3$ (+19.4 kcal mol$^{-1}$), (C$_6$F$_5$)$_3$P/B(C$_6$F$_5$)$_3$ (+20.1 kcal mol$^{-1}$), $t$Bu$_3$P/B(C$_6$H$_2$Me$_3$)$_3$ (+30.6 kcal mol$^{-1}$), $o$-C$_6$H$_5$Me$_4$N(C$_6$H$_4$)BH$_2$(+19.8 kcal mol-1) and Ph$_2$N(C$_6$H$_4$)B(C$_6$F$_5$)$_2$ (+30.7 kcal mol$^{-1}$), while slightly positive or negative values have been found for the reactive

FLPs, such as Mes$_2$P(CH$_2$)$_2$B(C$_6$F$_5$)$_2$ (+7.7 kcal mol$^{-1}$), Mes$_2$P(CH$_2$)$_4$B(C$_6$F$_5$)$_2$ (+5.8 kcal mol$^{-1}$) and tBu$_3$P/B(C$_6$F$_5$)$_3$ (-11.9 kcal mol$^{-1}$).

Moreover, the results presented in **Figure 1** show that ΔG values are generally more negative when solvent effects are considered (both toluene and IL) compared to values in gas phase, indicating that H$_2$ activation by FLPs is favored by the solvents, which is similar to previous DFT calculations, of which a COSMO-RS calculation showed that solvation free energy contributions from toluene were around -10 kcal mol$^{-1}$ [21]. Interestingly, we find that IL shows larger solvation free energy contributions than that of toluene, i.e, the solvation free energies of ILs are computed to be ranging from -10 kcal mol$^{-1}$ to -30 kcal mol$^{-1}$, while that of toluene ranges from -1 kcal mol$^{-1}$ to -10 kcal mol$^{-1}$. Taking inter-molecular FLPs as examples, four FLPs have been found showing large positive ΔG values in the gas phase, which are (C$_6$H$_2$Me$_3$)$_3$P/BPh$_3$, Tmp/BPh$_3$, P(C$_6$F$_5$)$_3$/B(C$_6$F$_5$)$_3$ and tBu$_3$P/B(C$_6$H$_2$Me$_3$)$_3$, having computed ΔG in gas phase being +20.3, +20.2, +23.6 and +34.6 kcal mol$^{-1}$, respectively. After considering solvent effects from toluene, the ΔG becomes +19.4, +16.7, +20.1 and +30.6 kcal mol$^{-1}$, respectively. These values continually become smaller (less positive) when solvent effects of IL were taken into account, which are +12.9, +10.6, +12.8 and +17.1 kcal mol$^{-1}$, respectively. The computed toluene solvent-phase ΔG of the rest five inter-molecular FLPs are -11.9, -11.1, -2.5, -11.7 and -21.1 kcal mol$^{-1}$, respectively. Again, these values become more negative if the IL is presented as the solvent in the DFT calculations, which are -16.5, -15.6, -9.5, -15.9 and -28.3 kcal mol$^{-1}$, respectively. Moreover, the results show that the solvent effects are also dependent on the types of FLPs. Taking ILs as example, the solvation free energies of inter-molecular FLPs are general smaller than that of intra-molecular FLPs. For inter-molecular FLPs, the smallest solvation free energy is from (C$_6$H$_2$Me$_3$)$_3$P/BPh$_3$, which is -7.4 kcal mol$^{-1}$, while the largest one is from tBu$_3$P/B(C$_6$H$_2$Me$_3$)$_3$, which is -17.5 kcal mol$^{-1}$. For intra-molecular FLPs, the smallest solvation free energy is from (C$_6$H$_2$Me$_3$)$_2$P(C$_6$F$_4$)B(C$_6$F$_5$)$_2$, which is –12.7 kcal mol$^{-1}$, while the largest one reaches –29.50 kcal mol$^{-1}$ in the case of o-C$_6$H$_5$Me$_4$N(C$_6$H$_4$)BH$_2$. Note that we have not found any inter-molecular FLPs changes the sign of ΔG values after including the solvation free energies. However, this is not the case for the intra-molecular FLPs. There is one intra-molecular FLPs changing the sign of ΔG values when DFT calculations were performed with the toluene as the solvent, which is $t\text{Bu}_2\text{P}(\text{C}_2\text{HMe})\text{B}(\text{C}_6\text{F}_5)_2$ (from +2.53 kcal mol$^{-1}$ to – 4.49 kcal mol$^{-1}$). Moreover, we find that four FLPs change the sign of ΔG values when ILs were considered as the solvents. They are $\text{Mes}_2\text{P}(\text{CH}_2)_2\text{B}(\text{C}_6\text{F}_5)_2$, $\text{Mes}_2\text{P}(\text{CH}_2)_4\text{B}(\text{C}_6\text{F}_5)_2$, $t\text{Bu}_2\text{P}(\text{C}_6\text{F}_4)_2\text{B}(\text{C}_6\text{F}_5)_2$ and $\text{Mes}_2\text{P}(\text{C}_6\text{F}_4)_2\text{B}(\text{C}_6\text{F}_5)_2$, with the ΔG changing from +7.7, +9.4, +1.2 and +3.9 kcal mol$^{-1}$ in toluene to -8.5, -1.0, -9.9 and -6.2 kcal mol$^{-1}$, respectively. As a summary, the DFT calculations reveal that ILs have larger solvation free energy contributions compared to the traditional toluene solvent, and the contributions to the intra-molecular FLPs are generally larger than inter-molecular FLPs.

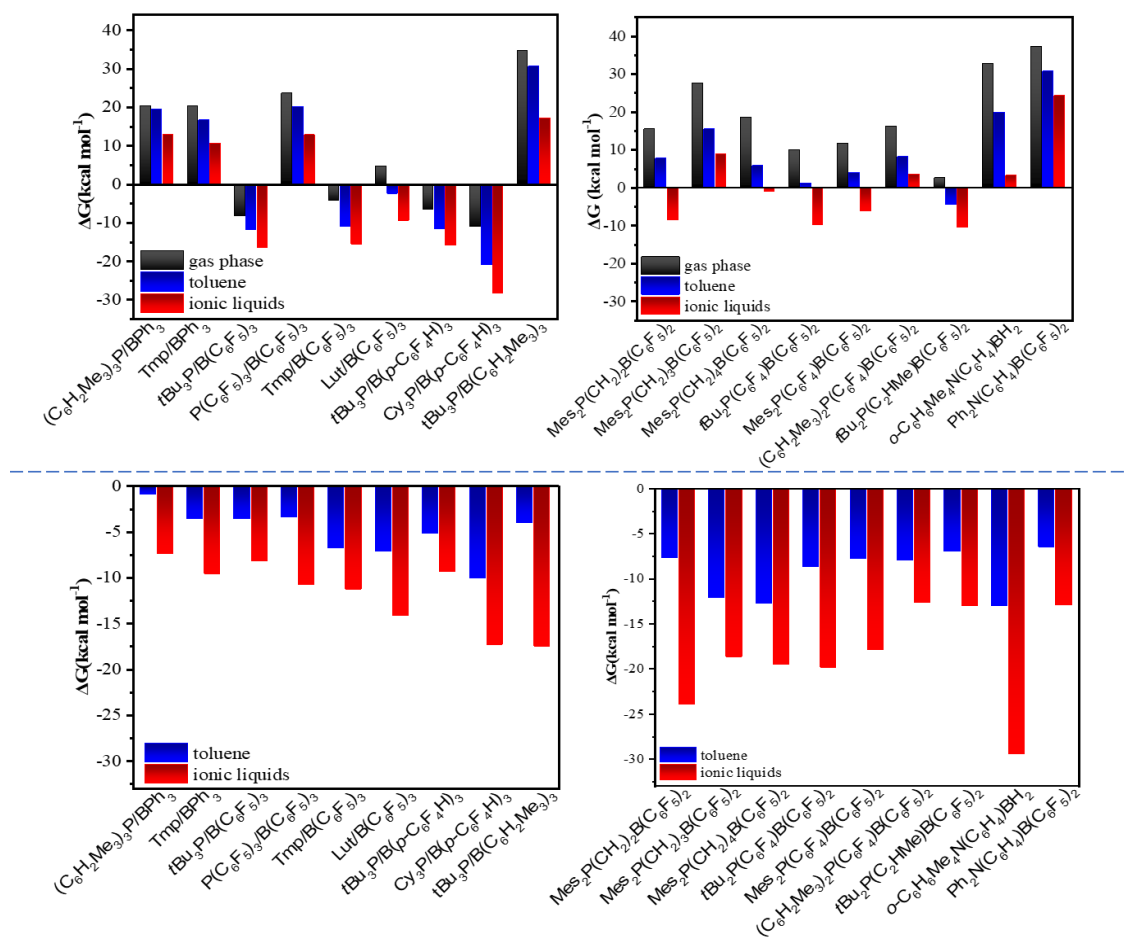

**Figure 1**. Top panel: computed total Gibbs free energies of inter-molecular (left) and intra-molecular (right) FLPs. Bottom panel: computed solvent effects for the total Gibbs free energies of inter-molecular (left) and intra-molecular (right) FLPs.

To gain deep insight into the solvent effects on the thermodynamics for the H$_2$ activation by FLPs, we employed a ΔG partitioning scheme which was proposed by

Pápai and co-works[29]. Here, we simplified the overall ΔG by the equation of ΔG = $\Delta G_{hh}$ + $\Delta G_{pa}$ + $\Delta G_{ha}$ + $\Delta G_{stab}$, where $\Delta G_{hh}$ is the endothermic H-H cleavage process, $\Delta G_{pa}$ and $\Delta G_{ha}$ describe the strength of the LB (proton affinity) and LA (hydride affinity), respectively, and the last term $\Delta G_{stab}$ stand for the interactions between two charged species (i.e. [LBH]$^+$, and [LAH]$^-$). In the next contents, we will term by term discuss these individual contributions with the focus on the difference between toluene and ILs.

The first term of the thermodynamic cycle for $H_2$ activation FLPs is the H-H cleavage, where $H_2$ is spit into the H$^+$ and H$^-$ ($\Delta G_{hh}$: $H_2$ → H+ and H-). This term is constant for all FLPs, with a computed $\Delta G_{hh}$ being +290.8 kcal mol$^{-1}$ by employing toluene as the solvent in DFT calculations. Interestingly, such value decreases dramatically to be +194.8 kcal mol$^{-1}$ if ILs were employed as the solvent in the SMD method. As pointed out by previous the DFT calculations[29], there might be some errors induced by the applied methodology. However, the difference between the values in toluene and in ILs should remain a similar trend. Hence, we conclude that H-H bond gets thermodynamically easier to be broken in ILs than that in toluene.

The second term is the $\Delta G_{pa}$ (LB + H$^+$ → [LBH]$^+$), of which the Lewis bases (i.e. P/N) capture the H$^+$ through the interactions between the lone paired electrons of P/N and the empty orbital (σ) of H$^+$. The computed values of $\Delta G_{pa}$ for all FLPs in toluene and ILs are depicted in **Figure 2** (top panel). Generally, we can see that ILs have less negative vales compared to that of toluene. In other words, ILs show negative impact on the capture of H$^+$ by Lewis bases. On average, the difference between toluene and ILs is about -30 kcal mol$^{-1}$. For instance, the computed $\Delta G_{pa}$ are −187.6, -182.8 and -189.9 kcal mol$^{-1}$ for inter-molecular FLPs ($(C_6H_2Mes)_3P/BPh_3$, Tmp/BPh$_3$ and $tBu_3P/B(C_6F_5)_3$ in toluene, respectively, while they are -148.9, -151.0 and -154.3 kcal mol$^{-1}$ in ILs. Similarly, the computed $\Delta G_{pa}$ in toluene of intra-molecular FLPs Mes$_2$P(CH$_2$)$_2$B(C$_6$F$_5$)$_2$, Mes$_2$P(CH$_2$)$_3$B(C$_6$F$_5$)$_2$ and Mes$_2$P(CH$_2$)$_4$B(C$_6$F$_5$)$_2$ are -143.0, -171.7 and -178.7 kcal mol$^{-1}$, respectively, while they decrease to be −149.5, -71.3 and -145.1 kcal mol$^{-1}$ in ILs. Note that there is one intra-molecular FLPs with large difference, which is Mes$_2$P(CH$_2$)$_3$B(C$_6$F$_5$)$_2$. The computed $\Delta G_{pa}$ values in ILs is +100.4 kcal mol$^{-1}$ smaller than that computed in toluene. The third term contributing to ΔG is the hydride affinity ($\Delta G_{ha}$: LA + H$^-$ → [LAH]$^-$), which describes the strength of Lewis acids (i.e. B) to attract the H$^-$ through the interactions between the empty orbitals of B and lone paired electrons of H$^-$, and the computed values are shown in

**Figure 2** (bottom panel). The results show similar phenomenon to that of the $\Delta G_{pa}$, that is, ILs show negative impact on the H$^-$ capture by Lewis acids. For all studied FLPs, always less negative values have been found for $\Delta G_{ha}$ in ILs compared to that in toluene. For instance, the computed $\Delta G_{ha}$ in toluene for inter-molecular FLPs $(C_6H_2Mes)_3P/BPh_3$, Tmp/BPh$_3$ and $t$Bu$_3$P/B$(C_6F_5)_3$ are -57.7, -57.7, and -87.1 kcal mol$^{-1}$, respectively, while corresponding values in ILs are -35.1, -35.1, and -55.2 kcal mol$^{-1}$, respectively; and the $\Delta G_{ha}$ in toluene for intra-molecular FLPs Mes$_2$P(CH$_2$)$_2$B(C$_6$F$_5$)$_2$, Mes$_2$P(CH$_2$)$_3$B(C$_6$F$_5$)$_2$ and Mes$_2$P(CH$_2$)$_4$B(C$_6$F$_5$)$_2$ are -69.3, -64.9, and -71.3 kcal mol$^{-1}$, respectively, while corresponding values in ILs are -48.5, -31.8, and -47.0 kcal mol$^{-1}$, respectively. It is worth to point out that the computed $\Delta G_{pa}$ values are generally larger than the $\Delta G_{ha}$ values in both toluene and ILs solvent-phases. Specifically, the computed $\Delta G_{pa}$ values are about -180 kcal mol$^{-1}$ in toluene (ca. -140 kcal mol$^{-1}$ in ILs), while the computed $\Delta G_{ha}$ values are about -80 kcal mol$^{-1}$ in toluene (ca. -50 kcal mol$^{-1}$ in ILs). Note that a previous DFT study at a higher level of theory (SCS-MP2/cc-pVTZ//B97D/6-31G(d)) also found that $\Delta G_{pa}$ are about two-times larger than the $\Delta G_{ha}$[19]. This finding reveals that the strength of Lewis bases are more important for the overall thermodynamics for the H$_2$ activation by FLPs, which is consistent with a previous molecular dynamics (MD) study[44].

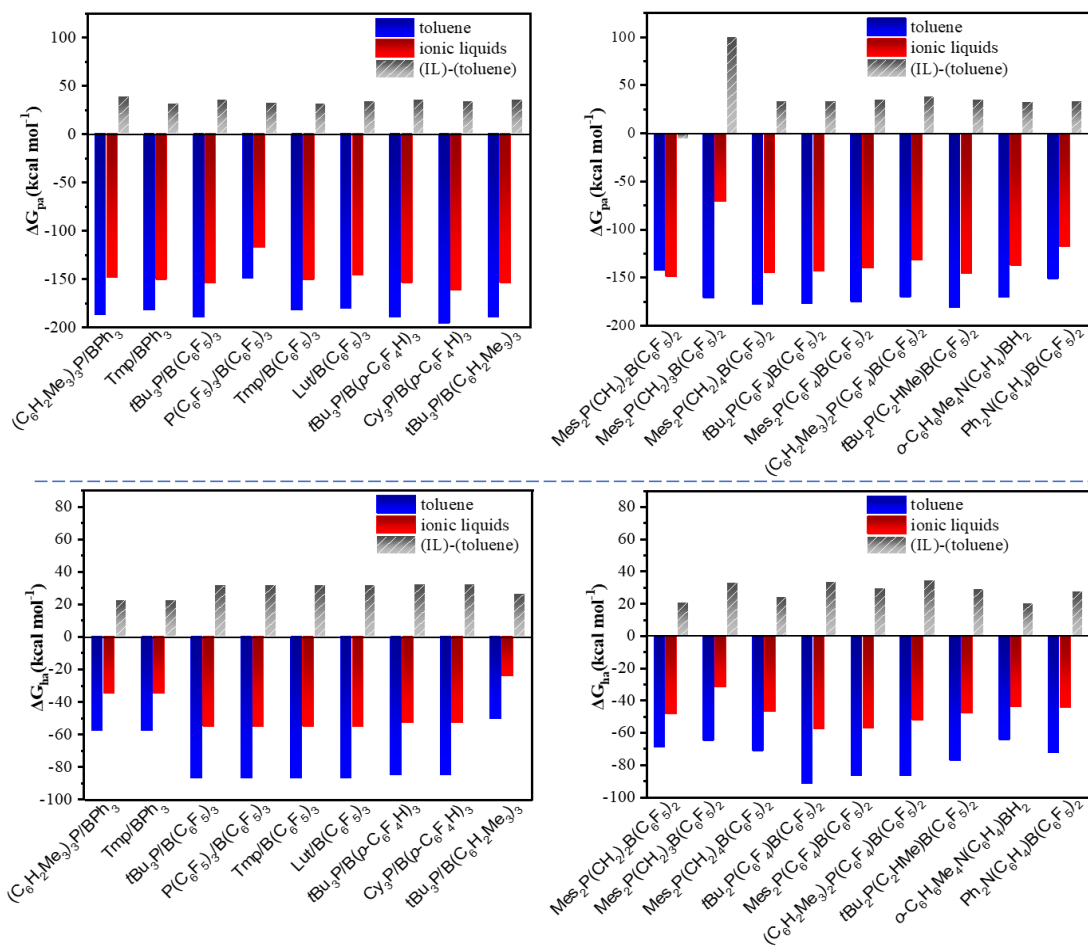

**Figure 2**. Top panel: computed the proton affinity ($\Delta G_{pa}$) of Lewis bases of the inter-molecular (left) and intra-molecular (right) FLPs. Bottom panel: computed the hydride affinity ($\Delta G_{ha}$) of Lewis acids of the inter-molecular (left) and intra-molecular (right) FLPs.

The final step of the H$_2$ activation by FLPs is the formation of zwitterionic pairs, with formulas of [LAH]$^-$[LBH]$^+$ and [$^-$HLA~LBH$^+$], respectively, for inter- and intra-molecular FLPs. The computed corresponding Gibbs free energies ($\Delta G_{stab}$) are summarized in **Figure 3**. When employing toluene as the solvent in the SMD calculations, the results show that the final zwitterionic products are strongly stabilized (negative values were found for all $\Delta G_{stab}$), and the intra-molecular FLPs generally have larger contribution from this term than the inter-molecular FLPs. The computed $\Delta G_{stab}$ of inter-molecular FLPs are around -28 kcal mol$^{-1}$ on average, i.e. the values of (C$_6$H$_2$Mes$_3$)P/BPh$_3$, Tmp/BPh$_3$, and $t$Bu$_3$P/B(C$_6$F$_5$)$_3$ are -26.1, -33.7 and -25.7 kcal mol$^{-1}$, respectively; while the computed $\Delta G_{stab}$ of intra-molecular FLPs range from ca. -20 kcal mol$^{-1}$ to -70 kcal mol$^{-1}$, i.e. the value of $t$Bu$_2$P(C$_6$F$_4$)B(C$_6$F$_5$)$_2$ is -20.7 kcal mol$^{-1}$ and the value of Mes$_2$P(CH$_2$)$_2$B(C$_6$F$_5$)$_2$ is -70.9 kcal mol$^{-1}$. When computed ΔG$_{stab}$ in ILs, the results show that obtained values in ILs are generally close to 0, indicating that ILs have negative impact on the stabilization process compared to toluene, of which the computed ΔG$_{stab}$ values are more than -20 kcal mol$^{-1}$. It is worth to point out that there is an exception having been found for intra-molecular FLPs, which is Mes$_2$P(CH$_2$)$_3$B(C$_6$F$_5$)$_3$. Its computed ΔG$_{stab}$ in ILs is larger (more negative) than that of in toluene, i.e. the ΔG$_{stab}$ in ILs is -82.8 kcal mol$^{-1}$, while the value in toluene is -38.8 kcal mol$^{-1}$.

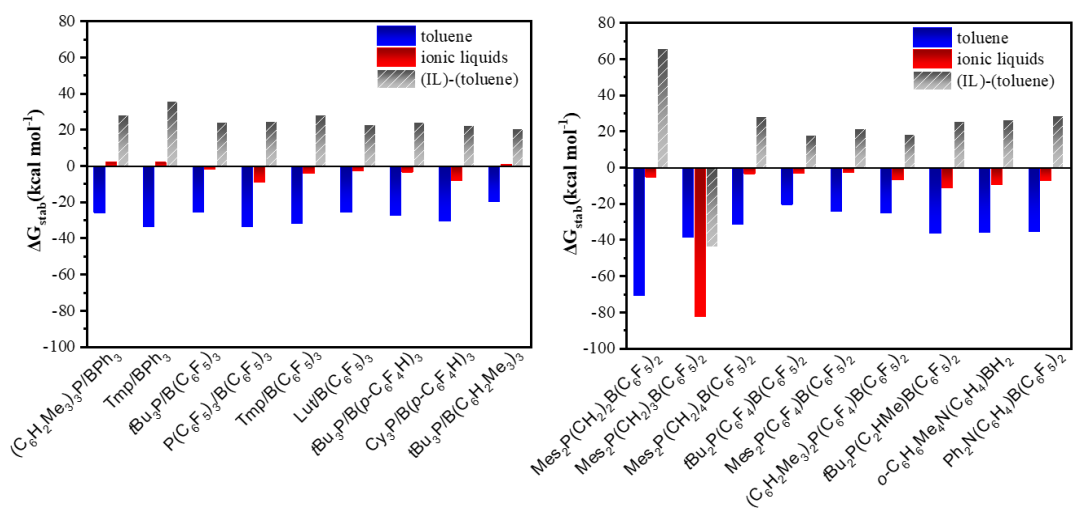

**Figure 3.** Computed stabilization Gibbs free energies (ΔG$_{stab}$) of inter-molecular (left) and intra-molecular (right) FLPs.

As a summary, the DFT calculations reveal that the ILs generally show positive impacts on the thermodynamics of H$_2$ activation by FLPs compared to traditional organic solvent (i.e. toluene), through increasing the solvation free energy contributions. The obtained findings could be summarized and explained from two aspects: 1) the computed ΔG in ILs are always more negative compared to the toluene when the products are charged species. In other words, ILs further stabilize ionic state products on the top of toluene. Such reactions include the overall H$_2$ activation, where zwitterionic products are formed ([LAH]$^-$[LBH]$^+$ or [$^-$HLA~LBH$^+$]), and the first elemental H-H cleavage reaction, where H$_2$ is spilt into a H$^+$ and a H$^-$. The computed ΔG in ILs are less negative for the elemental reactions of H$^+$/H$^-$ capture, and stabilization process compared to toluene. These results could be explained by the polarizabilities of the solvents, i.e. the dielectric constants, $\varepsilon$, are 2.37 and 24.85 for toluene and ILs, respectively; 2) the computed solvent-phase ΔG of intra-molecular FLPs are larger (more negative) than that of inter-molecular FLPs in the most cases. Such finding could be attributed to the charge separations and polarizabilities of the FLPs. As shown in **Table 1**, the charge difference between P/N and B atoms are quite similar for both inter- and intra-molecular FLPs, most of them are around 1.2 *e*. However, the distances between P/N and B atoms (*d*) of intra-molecular FLPs are about 2.0 Å longer than that in inter-molecular FLPs. The average *d* of inter-molecular FLPs are 3.5 Å, while average *d* of intra-molecular FLPs are 5.5 Å due to the bridge groups, i.e. $(CH_2)_n$, and $C_6F_4$. As such, the dipole moments of intra-molecular FLPs are generally larger than that of inter-molecular FLPs. The intra-molecular FLPs with the typical bridge groups of $(CH_2)_n$, and $C_6F_4$ have dipole moments being 60 and 80 $10^{-30}$ C·m, respectively, while average dipole moments of 50 $10^{-30}$ C·m have been found for the inter-molecular FLPs.

**Table 1.** The selected electronic properties of the zwitterionic products optimized in gas phase: the distance between Lewis acid and base centers, ***d***; the NPA charges of P/N atoms, ***q*<sub>P/N</sub>**; the NPA charges of B atoms, ***q*<sub>B</sub>**, and dipole moments, ***μ***.

|  | *d*/Å | $q_{P/N}$/e | $q_B$/e | $q_B$-$q_{P/N}$/e | $\mu/10^{-30}$C·m |
|---|---|---|---|---|---|
| **Inter-molecular FLPs** | | | | | |
| $BPh_3/(C_6H_2Me_3)_3P$ | 4.0 | 1.3 | 0.2 | -1.1 | 42.4 |
| $BPh_3$/Tmp | 3.3 | -0.8 | 0.2 | 0.9 | 37.5 |
| $B(C_6F_5)_3/tBu_3P$ | 4.2 | 1.4 | 0.1 | -1.3 | 57.1 |
| $B(C_6F_5)_3/(C_6F_5)_3P$ | 3.7 | 1.2 | 0.2 | -1.0 | 35.6 |
| $B(C_6F_5)_3$/Tmp | 3.6 | -0.7 | 0.1 | 0.8 | 50.8 |
| $B(C_6F_5)_3$/Lut | 3.5 | -0.5 | 0.1 | 0.6 | 48.2 |
| $B(p\text{-}C_6F_4H)_3/tBu_3P$ | 4.1 | 1.4 | 0.1 | -1.3 | 51.5 |
| $B(p\text{-}C_6F_4H)_3/Cy_3P$ | 4.1 | 1.4 | 0.1 | -1.3 | 48.1 |
| $B(C_6H_2Me_3)_3/tBu_3P$ | 4.4 | 1.4 | 0.1 | -1.2 | 51.0 |
| **Intra-molecular FLPs** | | | | | |
| $Mes_2P(CH_2)_2B(C_6F_5)_2$ | 4.2 | 1.3 | 0.1 | -1.2 | 65.8 |
| $Mes_2P(CH_2)_3B(C_6F_5)_2$ | 5.3 | 1.3 | 0.1 | -1.2 | 69.2 |
| $Mes_2P(CH_2)_4B(C_6F_5)_2$ | 5.9 | 1.3 | 0.1 | -1.2 | 67.9 |
| $tBu_2P(C_6F_4)B(C_6F_5)_2$ | 6.3 | 1.4 | 0.1 | -1.3 | 81.0 |
| $Mes_2P(C_6F_4)B(C_6F_5)_2$ | 6.2 | 1.4 | 0.1 | -1.3 | 86.8 |
| $(C_6H_2Me_3)_2P(C_6F_4)B(C_6F_5)_2$ | 6.2 | 1.4 | 0.1 | -1.3 | 81.3 |
| *t*Bu$_2$P(C$_2$HMe)B(C$_6$F$_5$)$_2$ | 4.3 | 1.4 | 0.1 | -1.3 | 53.6 |
| *o*-C$_6$H$_6$Me$_4$N(C$_6$H$_4$)BH$_2$ | 5.8 | -0.4 | -0.4 | 0.0 | 59.8 |
| Ph$_2$N(C$_6$H$_4$)B(C$_6$F$_5$)$_2$ | 3.3 | -0.4 | 0.0 | 0.5 | 42.3 |

**Conclusions**

In this study, we performed DFT calculations together with an implicit SDM solvent model to investigate the thermodynamics of H$_2$ activation by a series of inter- and intra-molecular FLPs in toluene and in ILs. The individual components and the solvation Gibbs free energy contributions have been examined, including H-H cleavage, proton/hydride attachment, and formation of zwitterionic products. The results show that ILs have more positive impact on the overall Gibbs free energies, that is, the computed overall ΔG in ILs are ~ -10 kcal mol$^{-1}$ more negative than that computed in toluene. Through the partitioning of the overall ΔG, we find that ILs favor the H-H cleavage elemental step more by ~100 kcal mol$^{-1}$, while disfavored ~30 kcal mol$^{-1}$ each for the elemental steps of proton/hydride attachment and zwitterionic stabilization. Moreover, the results show that these effects are strongly dependent on the type of FLPs, where intra-molecular FLPs are more effected compared to the inter-molecular FLPs by ~ -10 kcal mol$^{-1}$ for the overall ΔG. Based on the electronic structures analysis, i.e. distance between Lewis acid and base centers, NPA charges, and dipole moments of the final zwitterionic products, it is found that the DFT derived conclusions are somehow related to the polarizability: 1) the cation and anion consisted of ILs generally have larger polarizability compared to the neutral organic solvents; 2) intra-molecular FLPs products get more polarized due to the large charge separations of the localized partial positive and negative charges. With these theoretical investigations, we propose possibilities to construct ILs based FLP catalytic systems for H$_2$ activation and other important applications.

**Acknowledgement**

This work was financially supported by the National Natural Science Foundation of China (21978294, 21808224, 51901209, 21776281).